\documentclass[letterpaper,amsmath,amssymb]{revtex4}
\usepackage{graphicx}% Include figure files
\usepackage{dcolumn}% Align table columns on decimal point
\usepackage{bm}% bold math
\usepackage[usenames,dvipsnames]{color}
\usepackage{textcomp}

\newcommand{\betsgacl}{$\lambda$-(BETS)$_2$GaCl$_4$}
\newcommand{\betsfecl}{$\lambda$-(BETS)$_2$FeCl$_4$}
\newcommand{\kappabetsgacl}{$\kappa$-(BETS)$_2$GaCl$_4$}
\newcommand{\cecoin}{CeCoIn$_5$}

\newcommand{\etnhfour}{$\alpha$-(ET)$_2$NH$_4$(SCN)$_4$}

\newcommand{\drop}[1]{\relax}
\newcommand{\add}[1]{#1}
\newcommand{\note}[1]{\relax}

\begin{document}

%\preprint{APS/123-QED}

\title{\drop{Unusual} \add{Superconducting} phase diagram and FFLO signature in $\lambda$-(BETS)$_2$GaCl$_4$ from rf penetration depth measurements}% Force line breaks with \\

\author{William A.\ Coniglio}
% \altaffiliation[Also at ]{Physics Department, XYZ University.}%Lines break automatically or can be forced with \\
\author{Laurel E.\ Winter}%
\author{Kyuil Cho}%
\author{C.\,C.\ Agosta}%
\affiliation{Department of Physics, Clark University, Worcester, Massachusetts 01610}%
\author{B. Fravel}
\author{L.\,K.\,\ Montgomery}
\affiliation{Department of Chemistry, Indiana University, Bloomington, IN 47405}

\date{\today}% It is always \today, today,
             %  but any date may be explicitly specified

\begin{abstract}
We report the phase diagram of \betsgacl\ from rf penetration depth measurements with a tunnel diode oscillator in a pulsed magnetic field. \add{We examined four samples with 1100 field sweeps in a range of angles}\drop{Data from two samples is presented} with the magnetic field parallel and perpendicular to the conducting planes. In the parallel direction, $H_{c2}$ \add{appears to} include\drop{s} a tricritical point at 1.6 K and 10 T\drop{, and} \add{with a phase line that} increases to 11 T \drop{at} \add{as the temperature is decreased to} 500 mK. \drop{A}\add{The} second phase line \drop{develops from the tricritical point, forming} \add{forms} a clearly defined \add{high field low temperature} region satisfying \add{several of} the conditions of the Fulde-Ferrell-Larkin-Ovchinnikov (FFLO) state. \add{We show remarkably good fits of $H_{c2}$ to WHH in the reentrant $\alpha>1$, $\lambda_{so}=0$ regime.} We also note a sharp angle dependence of the phase diagram about the \add{field} parallel orientation that characterizes Pauli paramagnetic limiting\add{ and further supports the possibility of FFLO behavior}. \add{Unrelated to the FFLO study, at}\drop{At} fields and temperatures below $H_{c2}$ and $T_c$, we find \add{rich structure in the penetration depth data}\drop{unusual behavior} that we attribute to impurities at the surface altering the superconducting properties while maintaining the same crystallographic axes as $H_{c2}$.

\end{abstract}

% PACS, the Physics and Astronomy
                             % Classification Scheme.
%\keywords{Suggested keywords}%Use showkeys class option if keyword
                              %display desired
\maketitle
%{
%\renewcommand{\numberline}[1]{\relax}
%\tableofcontents
%}
\section{Introduction}
Organic conductors provide an excellent platform for researching unconventional superconductivity because they are clean low-dimensional materials whose critical temperatures and fields are accessible, yet their superconducting mechanisms are not yet fully understood. By developing the Tunnel Diode Oscillator \add{(TDO)} penetration depth measurement and pulsed field techniques, we aim to discover finer details in the phase diagram of the organics.

The cation BEDT-TSF (bis(ethylene-dithio)tetraselenafulvalene), usually called BETS, is closely related to the commonly studied BEDT-TTF (bis(ethylene-dithio)tetrathiafulvalene), which is shortened to ET, with the innermost four sulfur atoms of ET replaced by selenium to form BETS. The \betsgacl\ crystals are needle-shaped and crystalize alongside platelets of \kappabetsgacl, which do not superconduct. Layer spacing for \betsgacl\ is 18.58 \AA.~\cite{kobayashi1993} A very interesting relationship exists between the title compound and \betsfecl, which becomes an insulator at the title compound's critical temperature. \betsfecl\ also exhibits field-induced superconductivity, and may have an FFLO state of its own.\cite{uji_nature_2001,uji_prl_2006}

In a quasi-2d layered superconductor, the tunneling current between layers is small, but finite, so the layers act nearly independently, but maintain long-range coherence.\cite{singleton_coherence_prl_2002} If the vortices are able to fit between the layers, \add{the orbital limiting field increases dramatically, and superconductivity persists to much higher fields.}\drop{orbital quenching of superconductivity will be suppressed.}\cite{klemm_luther_1975,lee_chaikin02_prl} Barring other limits, when the Zeeman energy of a Cooper pair exceeds its binding energy at the Pauli paramagnetic limit, the pair will align parallel and the sample will no longer superconduct.\add{\cite{clogston62_prl,maki1966,hake1967} We have employed the theory of WHH\cite{werthamer_helfand_1966,collings_ti_book} to model $H_{c2}(T)$, taking into account both orbital and Pauli paramagnetic limiting.}

Paramagnetically limited materials present interesting possibilities for novel superconducting behavior. In a sample where the orbital limit\drop{ing has been suppressed} \add{is significantly larger than the Pauli paramagnetic limit}, a large magnetic field may cause the ground state wave function to become spatially modulated. In this case, the Cooper pairs acquire a non-zero linear momentum. This inhomogeneous superconducting state is known as the Fulde-Ferrell-Larkin-Ovchinnikov (FFLO or LOFF) state after the two groups who postulated it independently in 1964.\cite{ff_1964,lo_1965} \add{There have been a number of reports of FFLO behavior in various samples, although none have been definitive. Even so, the organic superconductors remain some of the ripest ground for the search. In the title compound, thermal conductivity measurements\cite{tanatar_ishiguro_prb_2002} show a cusp at the appropriate location in the phase diagram. There are also reports in other organics\cite{singleton_jpcm_2000,braunen_poster_2007,lortz_wang07_prl,cho2009,bergk_wosnitza_2010}, including some using the same technique we use here. Notably, \etnhfour\ does not exhibit any FFLO phase.\cite{coffey_nhfour_2010} In the heavy fermion superconductor} \cecoin, specific heat measurements\cite{radovan_fortune03_nature, bianchimovshovich03} \drop{identified the} \add{suggested} FFLO behavior, and penetration depth studies\cite{martin_agosta05_prb} followed\add{. Ref.~\cite{aperis_prl_2010} argues for a competition between ordered states that results in a spin density wave. Ref.~\cite{koutroulakis_prl_2010} suggests that the SDW may coexist with an FFLO state. The actual behavior is still unclear.}

\section{Experimental details}
Single-crystal samples of \betsgacl\ were grown using an electrochemical technique\cite{kobayashi1993,montgomery1993} in 1995--6 and stored \add{for 13 years}. We will follow the crystallographic convention that the long axis of the needle grows in the c-direction, and the best conduction occurs in the a-c plane with layers stacked along the b-axis. Samples A and C measured 0.1 mm each along the a- and b-axes and 0.3 mm along the c-axis. Sample~D had the same proportions in a slightly larger size. Sample B was slightly wider than A and C along the a-axis\drop{, but was otherwise almost identical.}\add{, electrically different from the other three, and does not agree with the literature on \betsgacl. It appears to superconduct, but at $T_c=4.25$ K and $H_{c2}=9.3$ T. We measured Sample~D ($T_c=5.1$ K) in a slightly modified experiment and obtained a limited amount of data, all of which was consistent with Samples A and C.}

\add{Penetration depth studies of \betsgacl\ have been difficult in the past due to the small size of the samples. We redesigned portions of the tunnel diode oscillator circuit to make effective use of standing waves that appear when using high frequencies.} We mounted the samples on a 1-axis rotator with $0.1^\circ$ resolution and ensuring precise alignment by pulsing the field up and down at a range of angles 0.23$^\circ$ apart about the parallel orientation. We oriented the sample on the rotator to apply the magnetic field along the b- or c-axes (Samples A and C) and along the b- or a-axes (Samples B and C). We ran Sample C twice, with fields parallel to the a- and c-axes, and noticed no significant differences. Orientation in the a-c plane was set upon sample mounting within a tolerance of $\pm15^\circ$. The magnetic field of the oscillator was always along the b-axis, so that currents are excited within the planes of the conductor.

We performed the experiments in a $^3$He cryostat inside the 40 T \add{medium} pulse\drop{d} magnet at Clark University with a rise-time of 27 ms. By using either $^3$He or $^4$He in the cryostat, we were able to take data from 367 mK to 4.3 K in a liquid environment\add{.} \drop{and above 3 K in gas.}

\begin{figure}\includegraphics[width=\textwidth]{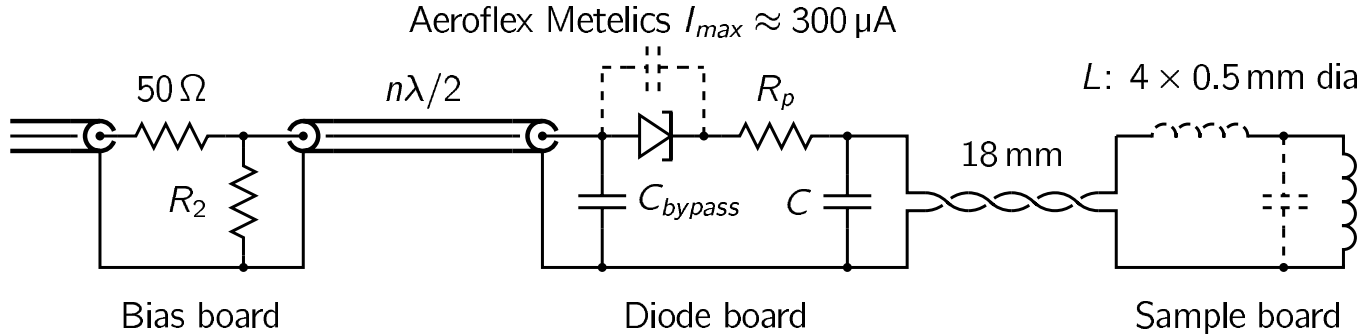}\caption{\label{fig:tdocircuit}Tunnel diode oscillator circuit used in 390 MHz examination of the organic superconductor \betsgacl. $R_2=166~\Omega$, $C_{bypass}=471$~pF, $R_p=100~\Omega$, $C=5$~pF. The length of the tuned coax cable is 511~mm.}\end{figure}
To measure the in-plane penetration depth, we designed a Tunnel Diode Oscillator\cite{vandegrift1975,coffey2000,gevorgian_rsi_1998} shown in Fig.~\ref{fig:tdocircuit} and described in \cite{coniglio_tdorsi}, to operate at 390 MHz with the sample contained in a 4-turn 0.5 mm diameter coil. For maximum sensitivity, the oscillator circuitry was placed next to the rotator and connected to the sample coil with 18 mm of twisted pair. We reduced the amount of heat generated near the sample by moving our biasing and matching resistor network above the $^3$He liquid level and used a length-tuned coaxial cable to connect the oscillator to the matching network. We achieved minimal noise in pulsed field despite locating the diode in the region of greatest field by using rf techniques such as impedance matching the signal into the 50 $\Omega$ coax leading up the probe, regulating the supply, and the use of ferrite chokes around coax terminations throughout. \add{With the LC oscillator free running at 390 MHz, our pre-pulse data gave a frequency measurement with a scatter of about $\pm200$ Hz, corresponding to an independent 0.5 ppm measurement every 13 \textmu s. The noise was about $10^5$ times smaller than the total frequency shift from the superconducting to normal state.}

Our RF receiver used a single-conversion to 5 MHz. The radio environment between 390 and 400 MHz was quiet, and conversion images were minimal. The conversion introduces a minus sign into the receiver frequency, and a graph of received frequency vs.\ magnetic field may be interpreted as a penetration depth vs.\ magnetic field, although it is difficult to obtain absolute units for the penetration depth.\cite{schawlow_devlin_1959,prozorov_giannetta_2006} We digitized the signal at 20 MS/s for computational frequency demodulation and analysis. \cite{coniglio_tdorsi}

\add{\subsection{Definition of $H_{c2}$}
\begin{figure}
\begin{center}
\includegraphics{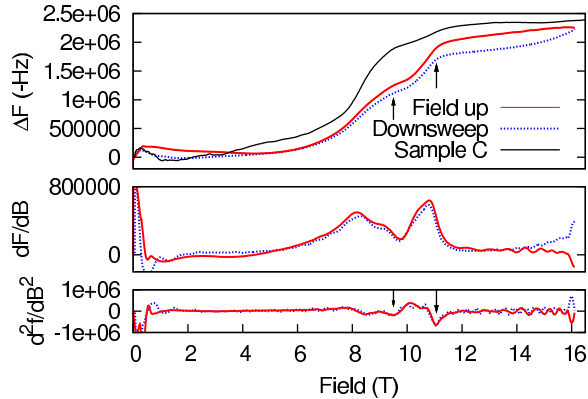}
\end{center}
\caption{\label{fig:exampletrace}(Color online) Example frequency vs.\ field \add{at 512~mK} with two derivatives shown. At low field, the sample is in the superconducting state. As the field increases, Cooper pairs are destroyed, and the London penetration depth $\lambda_L$ increases. \add{At 9.5~T (lower arrow), the sample enters the inhomogeneous FFLO state. As the magnetic field increases further, the inhomogeneous state can no longer support the magnetic field, and the sample transitions to the normal metallic state. (upper arrow)} When the penetration depth is larger than $\delta$, the metallic skin depth, the flat normal state dominates the RF response. \add{Also shown is a field sweep at the same temperature from Sample~C, which exhibits transitions at identical fields.}}
\end{figure}
There is inherent ambiguity in the definition of $H_{c2}$ because the transition is often broad. We favor identifying the critical field using minima in $d^2f/dB^2$, an approach that is well-defined, highly localized, requires little to no human judgement, and produces self-consistent results over many trials. We use the exact same criteria (minima in $d^2f/dB^2$) to identify the Vortex SC-FFLO, FFLO-Normal, and Vortex SC-Normal phase transitions. Inspection of the top plot in Fig.~\ref{fig:exampletrace} clearly show two complete transitions at low temperature in both samples -- the lower one from a vortex superconductor to FFLO, and the upper one from FFLO to a normal metal.

In addition, we have found that the critical temperature we determine by an analogous method (minima in $d^2f/dT^2$) agrees with the critical fields extrapolated to $T_c$. Our $T_c$ agrees with a specific heat study\cite{ishizaki_uozaki03}.

When in the superconducting state, the sample excludes RF energy from its center. The excluded volume lowers the inductance of the measurement coil, raising the oscillation frequency.\cite{prozorov_giannetta_2006} In the superconducting state, we measure the London penetration depth $(\lambda_L)$. In the normal state, the rf exclusion is governed by the metallic skin depth of the sample $(\delta)$, which depends on frequency and conductivity. From normal state resistivity measurements\cite{tanatar_ishiguro_prb_2002}, we expect skin depths from 25 to 80 \textmu m. We measure $H_{c2}$ at the point where the rapidly increasing penetration depth reaches the same size as the skin depth. We can represent that point very precisely by taking the second derivative of oscillation frequency as a function of field or temperature. We define $H_{c2}$ at the point where the concavity is a minimum, signifying a corner between metallic and superconducting behavior.}

\section{Results and discussion}
\drop{[removed entire results section, added new]}

We examined each sample in zero magnetic field and with the conducting layers both parallel and perpendicular to a magnetic field. Average $T_c$ was 4.93 K, in agreement with \cite{mielke_singleton_2001,ishizaki_uozaki03}. From our measurements, we have two reports to make. First, we will establish the phase diagram with the field perpendicular to the conducting layers using a full superconducting phase diagram measurement on a single sample and argue against the conclusion in \cite{mielke_singleton_2001} that \betsgacl\ exhibits a dimensional crossover at 1.9 K. Second, with the assistance of the WHH model, we will use our new penetration depth data with field parallel and nearly parallel to the conducting layers to identify the high field low temperature phase as a function of temperature, field, and angle, and affirm the suggestion of \cite{tanatar_ishiguro_prb_2002} that the phase may be due to the Fulde-Ferrell-Larkin-Ovchinnikov state. We observe that as the angle is rotated away from parallel to the planes, the transition between the vortex state and FFLO increases in field, while the introduction of vortices destabilizes the FFLO phase and causes its transition to the normal state to reduce in field. We attribute the vortex-FFLO behavior to an increase in spin-orbit scattering caused by vortices entering the conducting layers.

\subsection{Perpendicular phase diagram}
\begin{figure}
\begin{center}
\includegraphics{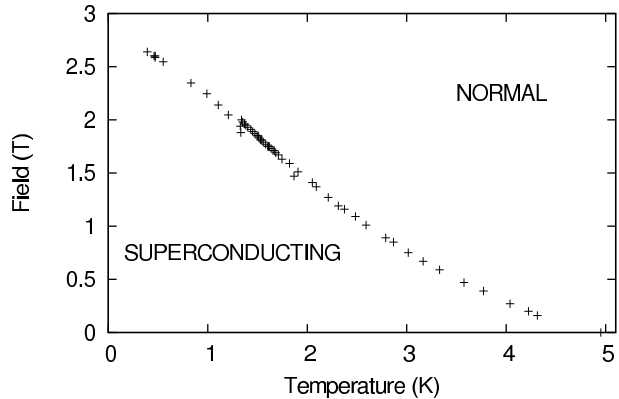}
\end{center}
\caption{\label{fig:pointsperpa}$H_{c2}^\perp(T)$. Each point represents an independent field sweep, with 53 sweeps total. Points shown are from the upsweep only, as vibration introduced extra noise into the downsweep. There was no hysteresis between sweeps. The slight difference in critical fields above and below 1.3 K reflects the use of liquid $^3$He or $^4$He. Because of the upward concavity above 2.5\,K, we were unable to realize an acceptable WHH fit in the perpendicular orientation.}
\end{figure}
Fig.~\ref{fig:pointsperpa} shows $H_{c2}^\perp$ as a function of temperature for a single sample. Comparing our phase diagram to Fig.~9 of \cite{mielke_singleton_2001}, we do not see the abrupt change in slope of $H_{c2}$ that they report at 1.9 K. Instead, we see a gradual increase in slope at about 3 K. Ref.~\cite{mielke_singleton_2001} suggested that at 1.9~K, the sample undergoes a dimensional crossover from a three-dimensional behavior to a quasi-two-dimensional one when the across-plane coherence length $\xi_\perp$ becomes smaller than the interlayer spacing. While we cannot rule out that explanation, we disagree with their conclusions regarding the dimensional crossover. The use of $H_{c2}^\perp$ to measure $\xi_\perp$ is not easily justified. Whether the sample is 2d or 3d, the in-plane size of a vortex depends on $\xi_\parallel$, and their cross-plane extent is the entire sample. Fields aligned parallel to the layers should be used to study the anisotropy and dimensionality, as we have done in Fig.~\ref{fig:pointsangle} and discuss in the section on angle dependence of $H_{c2}$.

We find two-dimensional behavior at both 1.35~K and 2.85~K. Since we do see a gradual change in slope near 3 K, we wish to keep open the question about the origin of this behavior. Due to the change in slope, WHH does not fit this data. WHH may not generally apply to the organics in the perpendicular orientation because samples are in the clean limit, where the mean free path is much greater than coherence length. Using our perpendicular orientation phase diagram and extrapolating to zero temperature, we find $\xi^\parallel_{T=0}=106 $\AA. 

\subsection{Parallel phase diagram and WHH fit}
\begin{figure}
\begin{center}
\includegraphics{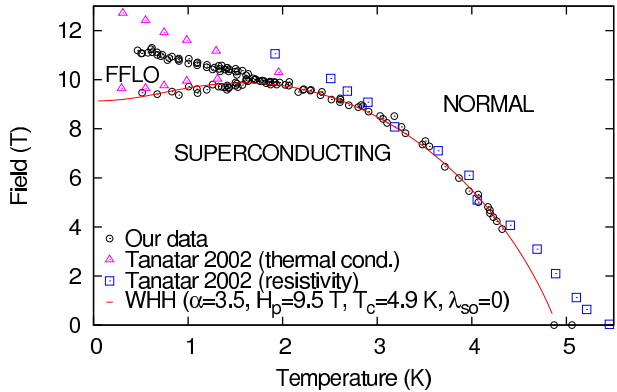}
\end{center}
\caption{\label{fig:pointspara}(Color online.) $H_{c2}^\parallel(T)$ from 116 magnet sweeps on Samples~A and~C. The two samples produced data that were indistinguishable from one another, and we plot them together here. We see no hysteresis in critical fields between the up and down sweeps. Below 1.7 K, a second minimum in the $d^2f/dH^2$ representation appears between 9 and 10 Tesla, which forms the lower branch on the phase diagram from the vortex superconducting state to the FFLO. Below 1.3 K in one of the samples, the Vortex SC-FFLO transition was partly masked by other features in the penetration depth, which we discuss later. Also shown are our WHH fit and the transport data from~\cite{tanatar_ishiguro_prb_2002}.}
\end{figure}
The phase at high field and low temperature has been ascribed to the FFLO state.\cite{tanatar_ishiguro_prb_2002} Our study has produced $H_{c2}$ and $H_{FFLO}$ values with significantly more detail than the previous work. Points from both measurement techniques of \cite{tanatar_ishiguro_prb_2002} and the present work appear in Fig.~\ref{fig:pointspara}.

We will now compare our data to the theory of Werthamer, Helfand, and Hohenberg\cite{werthamer_helfand_1966,collings_ti_book}. Starting from fixed $T_c=4.9$ K and no spin-orbit scattering ($\lambda_{so}=0$), we use the low temperature data to tune $H_p=9.5$~T and high temperature data to tune
\begin{equation}H_{c2}^0=0.692T_c\left(\frac{dH_{c2}}{dT}\right)_{T_c}=\alpha \frac{H_p}{\sqrt{2}}=23.5 \textnormal{~Tesla}\,.\end{equation}
From $H_{c2}^0$ and $H_p$, we find $\alpha=3.5$ and produce the fit shown in Fig.~\ref{fig:pointspara}. Remarkably, the fit reproduces the maximum $H_{c2}$ at 1.6 K, in excellent agreement with both the present work and the interpolated value in \cite{tanatar_ishiguro_prb_2002}. For comparison with \cite{agosta_martin06_jopcs}, using the Dingle temperature from \cite{mielke_singleton_2001} (whose samples came from the same batch as ours) and our $\xi^\parallel_{T=0}=106$~\AA, we find $3.5<r<12$. Using only data from \cite{mielke_singleton_2001}, $r$ might be 10\% lower.

Despite the good fit, we pay careful attention to the intended purpose of the WHH model: First, WHH was designed for bulk superconductivity, and no consideration is made for the effects of layers, particularly the orbital lock-in effect. We considered the theory of Klemm, Luther, and Beasley\cite{klemm_luther_1975}, which does examine the effect of layers. At low temperatures in the parallel orientation, the behavior of KLB and WHH is identical, and WHH require fewer parameters. While we continue to examine the application of KLB in hopes of modeling the microscopic behavior and tilted fields more carefully, the present parallel orientation data appears to be modeled well as a bulk-2d anisotropic superconductor with $T_c=4.9$~K, $\alpha=3.5$, $H_p=9.5$~T, $\lambda_{so}=0$, and the addition of FFLO behavior at temperatures below the reentrant point. In particular, WHH reproduce the maximum in $H_{c2}$, which appears to be the tricritical point for the FFLO phase. According to Maki\cite{maki1966}, that local maximum is also the temperature below which the transition becomes first order. Determining the order of a transition from penetration depth data is difficult, and we cannot make any strong statements. We instead point out that the present data has a qualitatively different shape from \cecoin, which has a second order transition at the lower field (change in slope of $\lambda_L$ in \cite{martin_agosta05_prb}) and a first order transition at the higher field (sharp increase in $\lambda_L$ in \cite{martin_agosta05_prb}).

Second, the model is stated to apply in the dirty limit, which sounds as though it should not apply to the clean organic superconductors. With the field aligned exactly in the parallel orientation, cyclotron orbits prevent electrons from traveling great distances in most directions without encountering a layer boundary. The presence of layers scattering normal electrons causes their mean free path to become short, and the dirty limit becomes an acceptable approximation.

Third, WHH require that spin-orbit scattering is a small portion of the total scattering. Since we must set the spin-orbit scattering parameter ($\lambda_{so}$) to zero, we have satisfied this limitation of the model. We suspect the orbital lock-in effect prevents scattering by removing the vortices from the strongly conducting layers and away from (in direct space) the majority of the Cooper pairs. Support for the suppression of spin-orbit scattering in the parallel orientation comes from Figs.~\ref{fig:pointsangle} and~\ref{fig:tempanglefield}, where we observe the transition from a vortex superconductor to FFLO (9.7~T in Fig.~\ref{fig:pointsangle}) in the parallel orientation rising as the angle is rotated very slightly away from parallel, and vortices begin to cut through the layers. When vortices enter the conducting layers, spin-orbit scattering begins to occur, and the critical field bounding the vortex superconducting state rises. The introduction of spin-orbit scattering in the WHH model indeed causes an increase in $H_{c2}$, even after the reduction in $\alpha$ due to anisotropy is considered through Tinkham's 2d equation. Other effects due to the orbital lock-in effect are also possible and are under consideration. WHH should not apply when the fields are aligned far from from the parallel orientation, as the mean free path extends into the clean limit. The model does not fit our perpendicular data at all.

We have attempted preliminary fits to other superconductors with reported FFLO states. While the transition from vortex superconductivity to FFLO in the organics seems to fit WHH, \cecoin\cite{martin_agosta05_prb} cannot be fitted to a reentrant WHH model because the tricritcal point is too low in temperature. The difference suggests that the mechanism for the high field, low temperature state in \cecoin\ is different from the organics. The competition between ordered states explored in \cecoin\cite{aperis_prl_2010}\cite{koutroulakis_prl_2010} therefore does not discount the possibility of FFLO in the organics.

\subsection{$H_{c2}$ vs.\ angle}
\begin{figure}
\begin{center}
\includegraphics{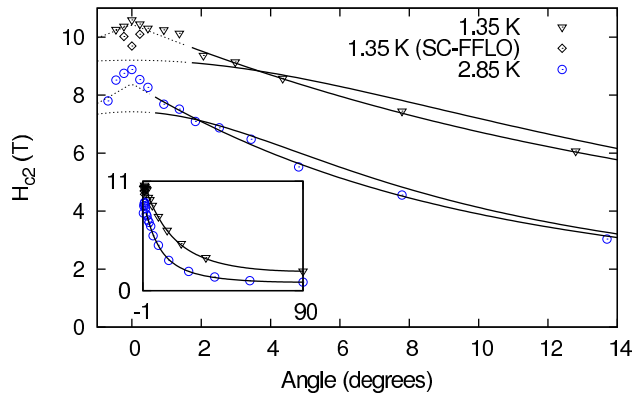}
\end{center}
\caption{\label{fig:pointsangle}(Color online.) $H_{c2}$ vs.\ angle from parallel orientation. G-L 3D and Tinkham 2d\cite{tinkhambook} curves from least-squares fits are shown with fit data extending to $90^\circ$. The 2d ones are sharply peaked, while the 3D are flat at $0^\circ$. The perpendicular end of the fits were fixed at measured values of $H_{c2}^\perp$. The 2d fits are both visually and statistically better at both temperatures, eliminating the possibility of a 2d/3d crossover between them. Parameters at 1.35 K are: 10.44 and 1.96 T, and at 2.85 K, 8.40 and 0.86 T. We avoid fitting the parallel data, because the fits do not account for the orbital lock-in effect, which appears as a nearly discontinuous increase in $H_{c2}$ within $2^\circ$ of parallel. Inset: Full range of data with 2d fit.}
\end{figure}
\begin{figure}
\begin{center}
\includegraphics{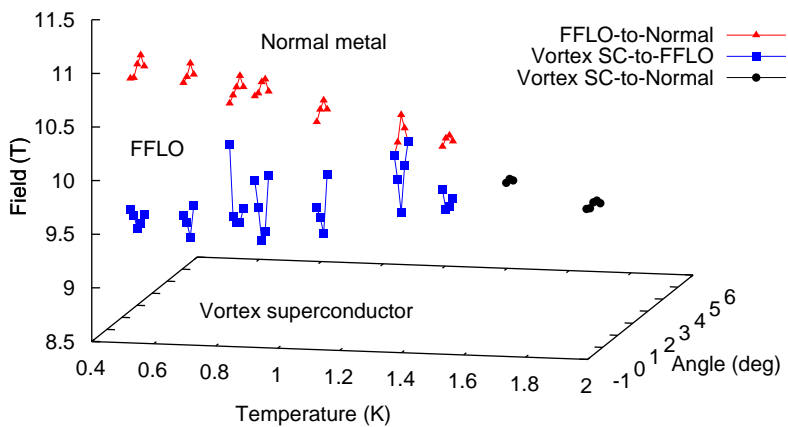}
\end{center}
\caption{\label{fig:tempanglefield}(Color online.) Here we illustrate the boundary of FFLO behavior as the temperature is lowered and field angle rotated away from parallel to the layers. Points connected by lines all share the same temperature. The dimension in and out of the page is the angle of the magnetic field from the layers. The reader may also consider it in 2d, where each group is at the x-axis temperature, and points connected by lines show the angular dependence of the critical fields with the apex at exactly parallel. The region between the blue and red markers is the FFLO state. At each temperature, the highest field FFLO-Normal transition and lowest field Vortex SC-FFLO transition both occur at the same angle (same field sweep, defined as $0^\circ$). As the angle is rotated in either direction away from $0^\circ$, the FFLO-Normal (red triangle) transition lowers in field, while the Vortex SC-FFLO one (blue squares) rises to meet it. See text for explanation of the rising behavior due to spin-orbit scattering. The angle sweep from Fig.~\ref{fig:pointsangle} appears here at 1.35~K.
}
\end{figure}
With Sample~A, we measured $H_{c2}$ vs.\ angle for a $91^\circ$ range from slightly before the parallel orientation to the perpendicular for temperatures 1.35 and 2.85 K. Resolution of the rotator was $0.12^\circ$, and our finest step size was $0.23^\circ$. The portion of both sweeps below $14^\circ$ is shown in Fig.~\ref{fig:pointsangle}. At both temperatures, even when the immediately parallel data is ignored, fitting to the Tinkham 2d form is superior to the G-L 3D one, eliminating the possibility of the 2d/3d crossover described in \cite{mielke_singleton_2001}. For small angles about the parallel, we see evidence of the orbital lock-in effect, as critical fields jump above the Tinkham prediction below a critical angle.

In order to ensure the exact parallel alignment of our sample to the magnetic field, we performed field sweeps at a range of angles near $0^\circ$. Critical fields as a function of angle and temperature appear in Fig.~\ref{fig:tempanglefield} as a 3d plot. The FFLO state is sideways pyramid-shaped pocket with its point at the tricritical point (1.7~K, $0^\circ$, 10.0~T). Of particular experimental interest, the critical field between the superconducting vortex state and FFLO rises as the angle is rotated away from parallel, even as $H_{c2}$ at 11~T falls.

\subsection{Features below $H_{c2}$}
\begin{figure}
\begin{center}
\includegraphics{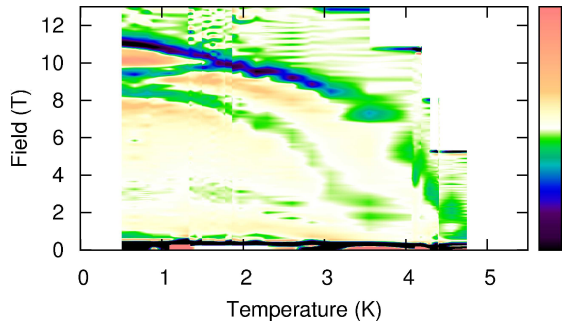}
\end{center}
\caption{\label{fig:parumap}(Color online.) Our concavity map from Sample~A indicates how easily we can identify portions of the phase diagram. The color key at right corresponds to the second derivative of oscillation frequency, $d^2f/dB^2$. Dark areas trace out the superconducting transition, in analogy with Fig.~\ref{fig:pointspara}. White areas are zero concavity. $H_{c2}$ is clearly visible as the darkest band from bottom-right to top-left. The transition between the vortex state and FFLO branches down from $H_{c2}$ at about 1.6 K and 10 T. The transition-like stripe at lower fields and temperatures is unexpected, and may reflect effects at the surface of the sample.}
\end{figure}
We have noticed features in our data below $H_{c2}$ and $T_c$. We attribute them to impurities at the surface of the sample, but their similarity to a superconducting transition interests us. One of them is visible in Fig.~\ref{fig:parumap} from 3.75 K to 8.5 T (green online), and also appears in Sample~C. A second only occurs in Sample~C and matches exactly with the only transition visible in Sample~B. The newer (and probably cleaner) Sample~D has no extra features. We have noted similar behavior in three samples of LiFeAs,\cite{cho_lifeas_2011} suggesting an effect of surface reconstruction or oxidation. The effect is visible as a deviation of penetration depth from the bottom of the diagonal lines in Fig.~1 of Ref.~\cite{cho_lifeas_2011}.

Lower features in both the title compound and LiFeAs appear very much the way a superconducting transition would, but do not appear to affect the bulk superconductivity in the material. They exhibit no hysteresis between the up and down sweep, often have $T_c$-like signatures in a zero-field temperature sweep, and have an angle dependence that follows the 3d G-L or 2d Tinkham forms, depending on the dimensionality of the sample.  The crystallographic axis is shared with the main transition at higher field, eliminating the possibility of a twinned crystal or a different morphology from consideration.

The coexistence of two forms of superconductivity that appear not to interact but share the same crystal structure is puzzling. If the surface of the samples exhibit different superconducting properties, the penetration depth measurement technique is sensitive to the behavior of one order parameter even if the other is fully superconducting, as the susceptibility of the sample will still change. Electrical conductivity, in particular, will not be sensitive to either the low field transition or the one between the ordinary superconducting state and FFLO one because the resistance will be zero below $H_{c2}$.

\drop{[end of added material]}

\section{Conclusion}
\drop{Using our phase diagram as a road map to interesting parts of the phase diagram, we suggest further measurements to explain each transition, particularly in the region we believe to be the FFLO state. We believe the greatest value experiment at this point would be a careful specific heat study on \betsgacl\ to focus on both the transitions below $H_{c2}$ and the possible FFLO state.} \note{We really need something here to suggest studies to identify the FFLO state. List compounds that might have it and press the community to positively identify the phenomenon.} \add{We show a phase diagram for \betsgacl\ with 116 field sweeps showing a high field, low temperature phase we attribute to the FFLO state. The angular dependence of the phase diagram is consistent with expected FFLO behavior. Our great point density allows us to reliably fit the ground state superconducting phase to the model by WHH. To our knowledge, this is the first example of experimental data that fits the reentrant WHH phase diagram. As no conclusive experiments have been done so far on the FFLO phase in the organics, it is difficult to prove that explanation. However, the body of evidence for the phase is growing, and it has been reported from a number of different measurement techniques. In addition to the ongoing phase diagram work by several groups, parallel-aligned high field studies of specific heat, neutron scattering, or STM, although difficult, would provide valuable confirmation or redirection of the research effort so far.}

\section{Acknowledgements}
\add{We would like to thank an anonymous reviewer, who prompted us to consider the WHH model.} We acknowledge support from the U.\,S. Department of Energy ER46214.

\bibliography{bibs/bets,bibs/organics,bibs/tdo,bibs/computation}{}

\end{document}